\documentclass[journal]{IEEEtran}

\usepackage[english]{babel}
\usepackage{mathptmx}

\usepackage{amsmath}
\usepackage{graphicx}
\usepackage[colorlinks=true, allcolors=blue]{hyperref}
\usepackage{float}
\usepackage[normalem]{ulem}
\usepackage{cite}
\usepackage{amsfonts}

\begin{document}

\title{Near-field orthogonality and cosine beams for near-field space division multiple access in 6G communications and beyond}

\author{{\IEEEauthorblockN{Sotiris Droulias, Giorgos Stratidakis, and Angeliki Alexiou,~\IEEEmembership{Member,~IEEE}} }
\thanks{The authors are with the Department of Digital Systems, University of Piraeus, Piraeus 18534, Greece (Corresponding author: Sotiris Droulias, e-mail: sdroulias@unipi.gr)}}

\maketitle

\begin{abstract}
Spatial division multiple access (SDMA), a powerful method routinely applied in multi-user multiple-input multiple-output (MIMO) communications, relies on the angular orthogonality of beams in the far field, to distinguish multiple users at different angles. 
%
%
%
Yet, with the gradual shift of wireless connectivity to the near-field of large radiating apertures, the applicability of classical SDMA becomes questionable. Therefore, to enable near-field multiple access, it is necessary to design beams that have the desired orthogonality in the near-field. %
In this work, we propose the concept of near-field space division multiple access (NF-SDMA), to enable SDMA in the near-field. 
%
%
We demonstrate analytically that the orthogonality of beams is preserved at any location of the receiver, from the near-field to the far-field of the transmitter. 
%
%
%
By judicious design, we select the family of cosine beams and we prove that they satisfy the orthogonality condition, offering a multitude of communication modes in the near-field. 
We demonstrate how the correlation of beams generated with uniform linear arrays (ULAs) is extended to uniform planar arrays (UPAs) in a straightforward and insightful manner. To test our analytical findings, we propagate the designed beams numerically, and we measure their orthogonality both at the transmitter and the receiver. We verify that the orthogonality of the proposed beams is successfully retrieved at a receiver that resides in the near-field of the transmitter, and is also robust to displacements of the receiver. Based on our findings, we propose codebook designs for NF-SDMA that are applicable for receivers with many elements and even with single antennas.
\end{abstract}

\section{Introduction}
\IEEEPARstart{W}{ireless} communications are nowadays shifting to higher frequencies, and the utilization of frequency bands, such as the mmWave and THz, is gradually becoming common ground for future wireless connectivity \cite{Tataria2021, Dai2024}. Higher frequencies involve shorter wavelengths and, because the size of everyday objects does not change along, several intriguing properties arise. For example, at shorter wavelength $\lambda$, a radiating element of size $D$ becomes electrically large, i.e. the ratio $D/\lambda$ increases. A direct consequence is that the Fraunhofer distance $z_F=2D^2/\lambda$ increases, and the near-to-far-field transition is transferred to larger distances, thus enabling operation of communications in the near-field of radiating apertures, such as reconfigurable intelligent surfaces (RISs) \cite{Pei2021, Tang2021, Singh2022a, Stratidakis2023, Droulias2024, Ahmed2024}, large antenna arrays (LAAs) and extremely large-scale antenna arrays (ELAA) equipped with thousands of antennas \cite{Arun2020, Dai2022, Ramezani2024}. \\
\indent In view of the high requirements of future wireless connectivity \cite{ITU2023}, the use of large radiating apertures is already being studied for manipulating the wavefront of the radiated wave, to acquire curvature beyond the typical far-field planar form \cite{Singh2023, Stratidakis2023, Droulias2024}. For example, the concept of beamfocusing is constantly gaining ground as a means to concentrate the power at controllable distances and small areas, a key element for energy efficient communications \cite{Huang2018, Ahsan2021, Tran2022, Yang2022, Zhang2022a, Li2023} and localization applications \cite{Wymeersch2020, He2022, Zhang2022b, Ma2023}. Recently, bending beams \cite{Singh2022b, Reddy2023} and beams with self-healing properties were proposed as an enabler for perceptive, resilient, and efficient networks \cite{Stratidakis2024}. \\ 
%
%
\indent Despite the various attempts to propose a near-field multiple access scheme, a strictly defined near-field orthogonality framework is still missing. \\
\subsection{Prior works}
\indent It is well-known that beam propagation in the far-field is equivalent to Fourier-transforming the beam at its starting location \cite{Droulias2022}. Hence, the properties of beams in the far-field can be accurately predicted, and this has enabled schemes, such as space division multiple access (SDMA) \cite{Roy1997} and beam division multiple access (BDMA) \cite{Sun2015, Dalela2018, Jiang2024}. These schemes rely on the angular orthogonality of beams, to ensure that at the maximum intensity of a certain beam, the interference from all other beams is minimized.  \\
\indent In the near-field, defining orthogonal beams in a similar manner becomes a challenging task. The reason is that the near-field of beams, generated with uniformly illuminated apertures, is characterized by spatial intensity oscillations that change rapidly for different propagation distances. Recently, in \cite{Dardari2021} it was demonstrated how to construct communication modes with focused beams, by taking advantage of the narrow extent of the focal area. The proposed orthogonality ensured minimum interference between beams, similar to how the maxima and minima of beams are treated in SDMA. In \cite{Dai2023a, Dai2023b} an orthogonality scheme for multi-antenna receivers was introduced for focused beams. It was shown that zero correlation between focused beams can be achieved only asymptotically, i.e. infinitely large arrays are required. In the latter works, the orthogonality was defined in terms of the steering vectors, which are equipped with the appropriate phase shifts needed to focus, rather than just steer beams. \\
\indent Evidently, there are various ways for defining orthogonality \cite{Aslan2021}. For beams in their far-field, the angular orthogonality \cite{Viberg1995} ensures that the peak of each beam coincides with the nulls of all the other beams \cite{Smith1988}. This is a direct consequence of the fact that the far-field is accurately predicted by the entries of the steering vectors; the latter also determinate the tilt of the beam's wavefront in the far-field, which becomes essentially flat with respect to the relatively small size of receiving apertures. In the near-field, however, the wavefront and intensity of the beam change locally with propagation distance, and it is not clear how the entries of the steering vectors can be associated with a definition of orthogonality. \\
%
%
%
\subsection{Our contributions}
In view of the various definitions of orthogonality, in this paper we define beam orthogonality for multi-beam antenna systems and we propose orthogonal near-field beams, which offer a multitude of communication modes for both multi-antenna and single-antenna receivers. The contributions are summarized as follows:
\begin{itemize}
    \item We introduce the concept of near-field SDMA (NF-SDMA) to bring into the near-field the classical far-field SDMA and its benefits, which are essential for novel near-field applications in future wireless communications.

    \item We prove analytically that, the orthogonality of beams is preserved at any location, from the near-field to the far-field of the transmitter.

    \item To design communication modes in the near-field, we select the class of cosine beams that offer strictly zero correlations for any array size. Our selection is inspired by the linearity of the phase shifts required to excite these beams, which is also a key aspect of the far-field SDMA orthogonality.

    \item We derive analytically the correlations of cosine beams and we provide analytical expressions for the orthogonality condition in ULAs and UPAs.

    \item We prove theoretically that there is a countable set of mutually orthogonal beams, all propagating along the same direction and converging at different distances from the transmitter. We also prove that the same conclusion holds for beams propagating along different directions and converging at the same distance from the transmitter. Additionally, we demonstrate that there is a multitude of beamsets involving beams at different angles and different convergence distances, where all beams are mutually orthogonal, providing an ideal means for NF-SDMA in both the angular and distance domain. 

    \item We numerically verify for different array sizes that the beam orthogonality of the proposed beams is successfully retrieved at a receiver that resides in the near-field of the transmitter, and is also robust to displacements of the receiver. 
   
    \item Based on our findings, we propose codebook designs for NF-SDMA that are applicable for receivers with many elements and even with single antennas.    
\end{itemize}
%
%
%
%
%
\subsection{Organization and notations}
The remainder of this paper is organized as follows. 
In Section \ref{sec:BeamOrthogonality} we define the condition for beam orthogonality.
In Section \ref{sec:CosineBeamsULAs} we introduce the family of cosine beams and we study their correlations for ULAs.
In Section \ref{sec:CosineBeamsUPAs} we extend the analysis to UPAs.
In Section \ref{sec:CodebookA} we propose a codebook design for NF-SDMA for multi-antenna receivers. 
In Section \ref{sec:CodebookB} we propose a codebook design for NF-SDMA for single-antenna receivers. 
In Section \ref{sec:Conclusion} we summarize the concluding remarks of this work.
\\\\
Notations: In this paper $\mathbb{Z}$ denotes the set of integers. ${\left[\cdot\right]}^T$ and ${\left[\cdot\right]}^H$ denote the transpose and conjugate-transpose operations, respectively; ${\left|\cdot\right|}$ denotes the norm of its complex argument and the asterisk $*$ denotes the complex conjugate. 
\section{Beam orthogonality} \label{sec:BeamOrthogonality}
%
%
%
\begin{figure}[t!]
\centering
		\includegraphics[width=1\linewidth]{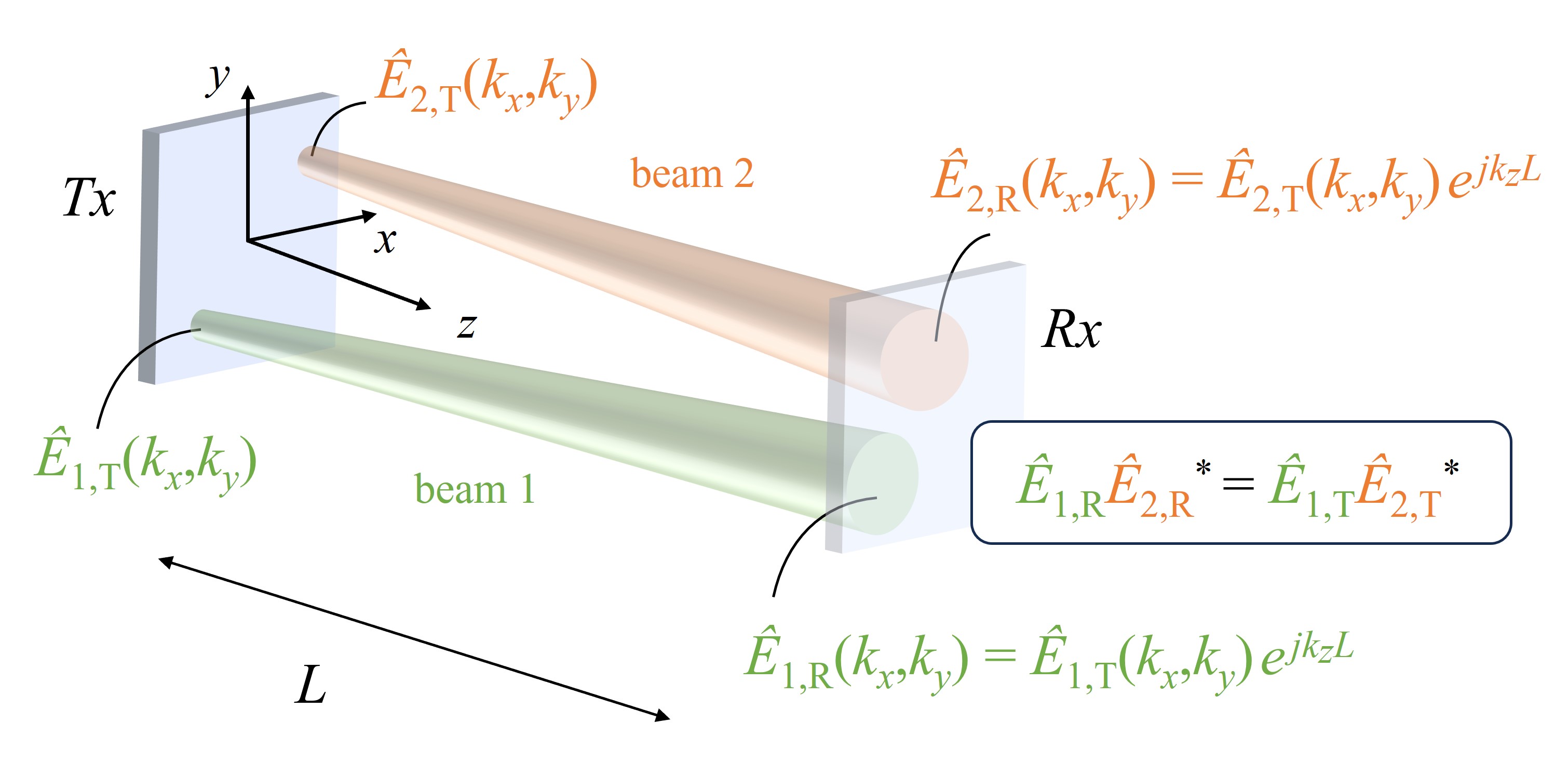}
	\caption{Invariance of beam correlations from the near-field to the far-field, explained in $k$-space. Schematic illustration of two beams propagating from the transmitter (Tx) to the receiver (Rx). The footprints of beams 1 and 2 at the receiver, $\hat{E}_{1,R}$ and $\hat{E}_{2,R}$,  are expressed in terms of their respective footprints at the transmitter, $\hat{E}_{1,T}$ and $\hat{E}_{2,T}$. The correlation of the two beams remains invariant in $k$-space and, consequently, in real space.}
    	\label{fig:fig01}
\end{figure}
%
%
%
\noindent Let us consider a transmitter (Tx), such as a uniform planar array (UPA), centered at the origin of the coordinate system, with its plane extending on the $xy$-plane, as shown in Fig.\,\ref{fig:fig01}. A receiver (Rx) is located at distance $L$, with its surface also residing on the $xy$-plane. The Tx generates beams that propagate along the $z$-axis, the $E$-field of which at the observation point $\textbf{\textrm r}=$($x,y,z$) is $\textbf{\textrm E}(\textbf{\textrm r}) = \textbf{\textrm E}(x, y, z)$. The correlation of two such beams $\textbf{\textrm E}_1(x,y,z),\textbf{\textrm E}_2(x,y,z)$ on the transverse $xy$-plane is then written at distance $z=L$ as
\begin{align}
    C=\frac{\iint_{-\infty}^{+\infty} E_1(x,y;L)E_2^*(x,y;L)dxdy}{\sqrt{\iint_{-\infty}^{+\infty} |E_1(x,y;L)|^2dxdy}\sqrt{\iint_{-\infty}^{+\infty} |E_1(x,y;L)|^2dxdy}},
    \label{Eq:EqCstep1}
\end{align}
where $E_{1,2}$ are the magnitudes of the $E$-field vectors. According to Parseval’s identity \cite{Osgood2019}, the inner product of $E_1,E_2$ is given equivalently in $k$-space via the equality
\begin{align}
    \begin{split}
     \iint_{-\infty}^{+\infty} E_1(x,y;L)E_2^*(x,y;L)dxdy  = \\
      =\iint_{-\infty}^{+\infty} \hat{E}_1(k_x,k_y;L)\hat{E}_2^*(k_x,k_y;L)dk_xdk_y,
    \end{split}
    \label{Eq:EqCstep2}
\end{align}
where $x, y$ are the Cartesian transverse coordinates, $k_x, k_y$ the corresponding spatial frequencies, and the hats denote the waves in Fourier space. Using \eqref{Eq:EqCstep2}, the beam correlation function can be written in $k$-space as
\begin{align}
    C=\frac{\iint_{-\infty}^{+\infty} \hat{E}_1(k_x,k_y;L)\hat{E}_2^*(k_x,k_y;L)dk_xdk_y}{\sqrt{\iint_{-\infty}^{+\infty} |\hat{E}_1(k_x,k_y;L)|^2dk_xdk_y}\sqrt{\iint_{-\infty}^{+\infty} |\hat{E}_2(k_x,k_y;L)|^2dk_xdk_y}}.
    \label{Eq:EqCstep3}
\end{align}
At this point, we can take advantage of the fact that the Fourier spectrum of a beam at $z=L$, can be expressed in terms of its spectrum at $z=0$ as (see Appendix \ref{Sec:AppendixA} for details)
\begin{align}
    \hat{E}(k_x,k_y;L) = \hat{E}(k_x,k_y;0) e^{jk_zL},
    \label{Eq:Eq02}
\end{align}
where $k_z=\sqrt{k^2-k_x^2-k_y^2}$ and $k$ is the wavenumber. Using \eqref{Eq:Eq02}, the correlation function \eqref{Eq:EqCstep3} takes the form
\begin{align}
    C=\frac{\iint_{-\infty}^{+\infty} \hat{E}_1(k_x,k_y;0)\hat{E}_2^*(k_x,k_y;0)dk_xdk_y}{\sqrt{\iint_{-\infty}^{+\infty} |\hat{E}_1(k_x,k_y;0)|^2dk_xdk_y}\sqrt{\iint_{-\infty}^{+\infty} |\hat{E}_2(k_x,k_y;0)|^2dk_xdk_y}},
    \label{Eq:EqCstep4}
\end{align}
which is \textit{independent of the propagation distance}. Therefore, the correlation calculated at the Tx is preserved along the entire propagation path. This observation is of central importance, as it allows us to design beams with the desired correlations at the transmitter, guaranteeing that the correlation is also preserved at the receiver, regardless of whether the receiver is in the near- or far- field of the transmitter. Importantly, for near-field applications, where the beams are designed to be spatially localized (e.g. focused beams, non-diffracting beams) the beam is practically captured entirely by the receiver, and the integration limits in \eqref{Eq:EqCstep1} can be limited to the extent of the Rx surface. This is in contrast to the case of far-field beams, the spatial extend of which is much larger of typical antenna sizes. The correlation at the Tx is simply written as \\
\begin{align}
    C=\frac{\iint_{S_T} E_1(x,y;0)E_2^*(x,y;0)dxdy}{\sqrt{\iint_{S_T} |E_1(x,y;0)|^2dxdy}\sqrt{\iint_{S_T} |E_2(x,y;0)|^2dxdy}},
    \label{Eq:EqCstep5}
\end{align}
where $S_T$ is the Tx surface on which integration takes place. For arrays consisting of $N_x \times N_y$ elements, periodically arranged with periodicity $d_x$ and $d_y$ along the $x$ and $y$ directions, respectively, the Cartesian coordinates are discretized, $x=n_x d_x,y=n_yd_y$ ($n_x,n_y\in \mathbb{Z}$), and \eqref{Eq:EqCstep5} takes its discrete form, which reads
\begin{align}
    C=\frac{\sum_{n_x}\sum_{n_y} E_1(n_x,n_y)E_2^*(n_x,n_y)}{\sqrt{\sum_{n_x}\sum_{n_y} |E_1(n_x,n_y)|^2}\sqrt{\sum_{n_x}\sum_{n_y} |E_2(n_x,n_y)|^2}}.
    \label{Eq:EqCstep6}
\end{align}
Note that, for uniform amplitude $|E_1|=|E_2|=1$, the denominator becomes $N_x N_y$ and \eqref{Eq:EqCstep6} takes the frequently met form
\begin{align}
    C=\frac{1}{N_xN_y} \sum_{n_x}\sum_{n_y} E_1(n_x,n_y)E_2^*(n_x,n_y).
    \label{Eq:EqCstep7}
\end{align}
%
%
%
\indent Based on this general principle we can focus on the transmitter side to design beams with the desired orthogonality. In SDMA, the far-field orthogonality is associated with correlations of flat wavefronts at the Rx, which have the same (linear) phase gradient of the steering vectors that generate them. Hence, the SDMA orthogonality is associated with the linear phase shifts at the Tx. Inspired by this property, we select the family of cosine beams, the excitation of which requires linear phase shifts of the form $ k \beta |x|$, where $\beta$ is a tuning parameter. 
%
%
Following our design principle, next, we prove that cosine beams offer beamsets with strict orthogonality $C=0$, enabling a multitude of communication modes, and opening up vast possibilities for near-field multiple access schemes, such as the concept of near-field space division multiple access (NF-SDMA).
\section{Cosine beams for NF-SDMA} \label{sec:CosineBeamsULAs}
%
%
%
\begin{figure}[t!]
\centering
		\includegraphics[width=1\linewidth]{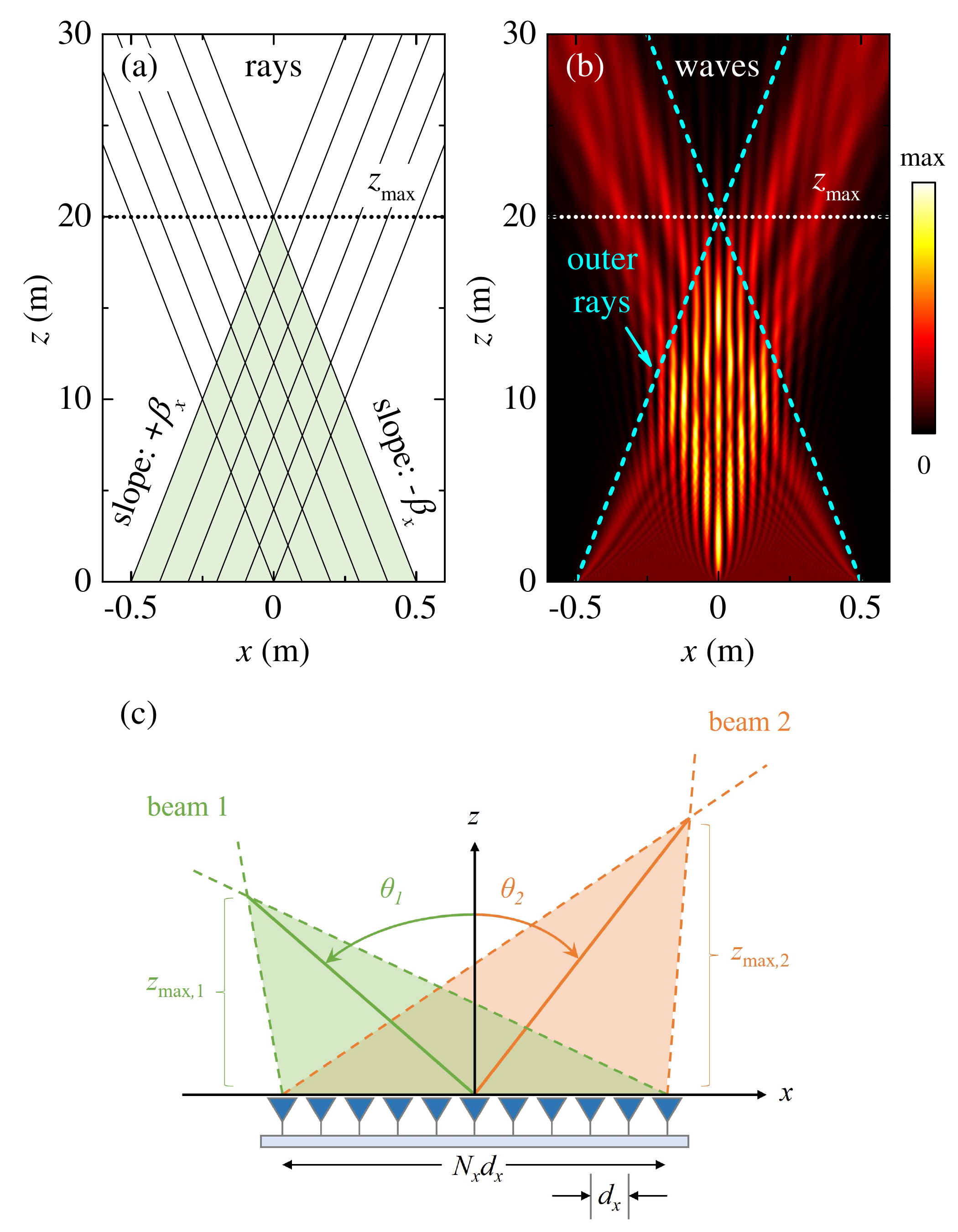}
	\caption{Cosine beams for NF-SDMA. (a) Ray representation. The shaded triangular region marks the beam content within the cosine beam range, the latter extending from the ULA up to distance $z_\mathrm{max}$. (b) Intensity of equivalent numerically propagated beam, using an array of size $N_x=1000$. The operation frequency is 150 GHz ($\lambda=2\,\mathrm{mm}$) and $d_x=\lambda/2$, here and throughout all the examples in this work. (c) System model for ULA communication systems, depicting two cosine beams with parameters ($\theta_1,z_\mathrm{max,1}$) and ($\theta_2,z_\mathrm{max,2}$), respectively. The shaded regions mark the beam content enclosed by the outer rays of each beam, as denoted in (a).}
    	\label{fig:fig02}
\end{figure}
%
%
\noindent Let us consider a large uniform linear array (ULA), centered at the origin of the coordinate system and extending on the $x$-axis. The ULA consists of $N_x$ elements with spacing $d_x$ and generates beams with tailored amplitude and phase characteristics that propagate on the $xz$-plane, as shown in Fig.\,\ref{fig:fig02}. The $E$-field profile at the input plane ($z=0$) has the form
\begin{align}
    E(x,0)=A(x) e^{j\phi(x)},
    \label{Eq:EqINPUTWAVE}
\end{align}
where $A$ is the amplitude and $\phi$ the phase. \\
%
%
\indent A cosine beam is a non-diffracting wave, formed by the superposition of two waves traveling at opposite oblique angles \cite{Jiang1997, Bencheikh2020}. A straightforward way to implement such beams is to introduce linear phase shifts with opposite slope around $x=0$, while keeping uniform amplitude ($A=1$). Hence, we may consider the following phase profile at the input plane
\begin{align}
    \phi(x)=k\sin{\theta}x-k\beta_x|x|,
    \label{Eq:EqIMPUTphase1Dcont}
\end{align}
where $\theta$ is the steering angle, i.e., a global angle, along which the entire beam is directed, and $\beta_x$ is a constant, associated with the slope of the two counter-propagating constituents of the beam. Note that, for $\beta_x=0$, \eqref{Eq:EqINPUTWAVE} accounts for conventional beam steering. The form of the $k\beta_x|x|$ term expresses the fact that parts of the beam that reside at opposite sides of the ULA (with respect to its center) have in-plane $k$-components with opposite signs. In the equivalent ray representation, this corresponds to rays with opposite slopes, namely $(1/k) (\partial \phi / \partial x) = \pm \beta_x$ with respect to the $z$-axis, as shown schematically in Fig.\,\ref{fig:fig02}(a). Due to the opposite slopes, the rays converge and overlap within a rhombic area, where a standing wave is formed. This area is enclosed by the outer rays (rays starting from the edges of the ULA), which intersect at a maximum distance $z_\mathrm{max}$, before they continue propagating individually towards opposite directions. Therefore, the parameter $\beta_x$ simply controls the beam convergence and, as we show in the Appendix \ref{Sec:AppendixB}, it is associated with the distance $z_\mathrm{max}$ as
\begin{align}
    \beta_x=\frac{N_x d_x}{2z_\mathrm{max}}.
    \label{Eq:EqBETA}
\end{align}
The distance $z_\mathrm{max}$ marks the cosine beam range, i.e. the extent of the standing wave, where the beam has non-diffracting properties, as shown in the ray representation in Fig.\,\ref{fig:fig02}(a) and also in the numerically propagated beam in Fig.\,\ref{fig:fig02}(b). The operation frequency is 150 GHz ($\lambda=2\,\mathrm{mm}$) and $d_x=\lambda/2$, here and throughout all the examples in this work. \\
\indent Note that cosine beams can also be excited as $E(x,0) = (1/2)(\exp(i k \beta_x x) + \exp(-i k \beta_x x)) = \cos(k \beta_x x)$ \cite{Bencheikh2020}. While this form also leads to two counter-propagating waves and the formation of a cosine beam, the input profile takes the form $A(x) = \cos( k \beta_x x),  \phi(x) = 0$, i.e. the beam is controlled via the amplitude of the array elements, rather than the phase. \\
\indent It is also important to note that the linearity of \eqref{Eq:EqIMPUTphase1Dcont} in $x$ is an advantage of cosine beams over other advanced beams, because it is usually easier to achieve linear than nonlinear phase shifts (e.g., quadratic, as required for focused beams), where phase oscillations can become very rapid, beyond the resolving capabilities of the array elements. Importantly, linear phase shifts are routinely applied in beamformers and RISs. Hence, the technology is well-known and it is straightforward to extend its use for the excitation of cosine beams.
\subsection{Beam correlations at the transmitter}
\noindent To calculate the orthogonality of cosine beams, we start with discretizing the continuous phase on the ULA grid, i.e. we express \eqref{Eq:EqIMPUTphase1Dcont} at points $x=n_xd_x$. The discretized version of the phase acquires the following form
\begin{align}
    \phi(n_x)=kd\sin{\theta}n_x-k d_x\beta_x|n_x|,
    \label{Eq:EqIMPUTphase1Ddisc}
\end{align}
where
\begin{align}
    n_x = -\frac{N_x-1}{2},-\frac{N_x-1}{2}+1,\dots,\frac{N_x-1}{2}.
    \label{Eq:EqNx}
\end{align}
The steering vector that imposes the phase profile \eqref{Eq:EqIMPUTphase1Ddisc} on the elements of the ULA, is simply given by
\begin{align}
\textbf{a}_L(\theta,\beta_x) = \frac{1}{\sqrt{N_x}}\left[e^{\phi(-\frac{N_x-1}{2})},e^{\phi(-\frac{N_x-1}{2}+1)},\dots, e^{\phi(\frac{N_x-1}{2})} \right]^T,
    \label{Eq:EqSV}
\end{align}
where the subscript $L$ denotes that it refers to a linear array. As we show in the Appendix \ref{Sec:AppendixC}, the correlation of two beams characterized by ($\theta_1,z_\mathrm{max,1}$) and ($\theta_2,z_\mathrm{max,2}$), respectively, can be equivalently expressed in terms of the steering vectors as
\begin{align}
    \begin{split}
      C_L(w_\theta,w_z)=|\textbf{a}_L(\theta_1,\beta_{x,1})^H \textbf{a}_L(\theta_2,\beta_{x,2})| = \\
      =\left |\frac{1}{N_x}\sum_{n_x} e^{j w_\theta n_x}e^{j w_z |n_x|} \right |,
    \end{split}
    \label{Eq:EqCL}
\end{align}
where 
\begin{subequations}
    \begin{align}
        &w_\theta=kd_x(\sin{\theta_1}-\sin{\theta_2}), \label{Eq:EqPAR1Da} \\
        &w_z=kd_x(\beta_{x,1}-\beta_{x,2})=kd_x\left(\frac{N_x d_x}{2z_\mathrm{max,1}}-\frac{N_x d_x}{2z_\mathrm{max,2}}\right),  \label{Eq:EqPAR1Db}
    \end{align}
\end{subequations}
express the angular detuning and convergence distance detuning, respectively, between the two beams. Two such beams are depicted schematically in Fig.\,\ref{fig:fig02}(c). Before investigating the general expression \eqref{Eq:EqCL}, we will first analyze two characteristic special cases, namely $w_\theta=0,w_z\neq0$ and $w_\theta\neq0,w_z=0$. \\
\subsubsection[Beams with u1 = u2]{Beams with $\theta_1=\theta_2$ and $z_\mathrm{max,1}\neq z_\mathrm{max,2}$}
\begin{figure}[t!]
\centering
    \includegraphics[width=1\linewidth]{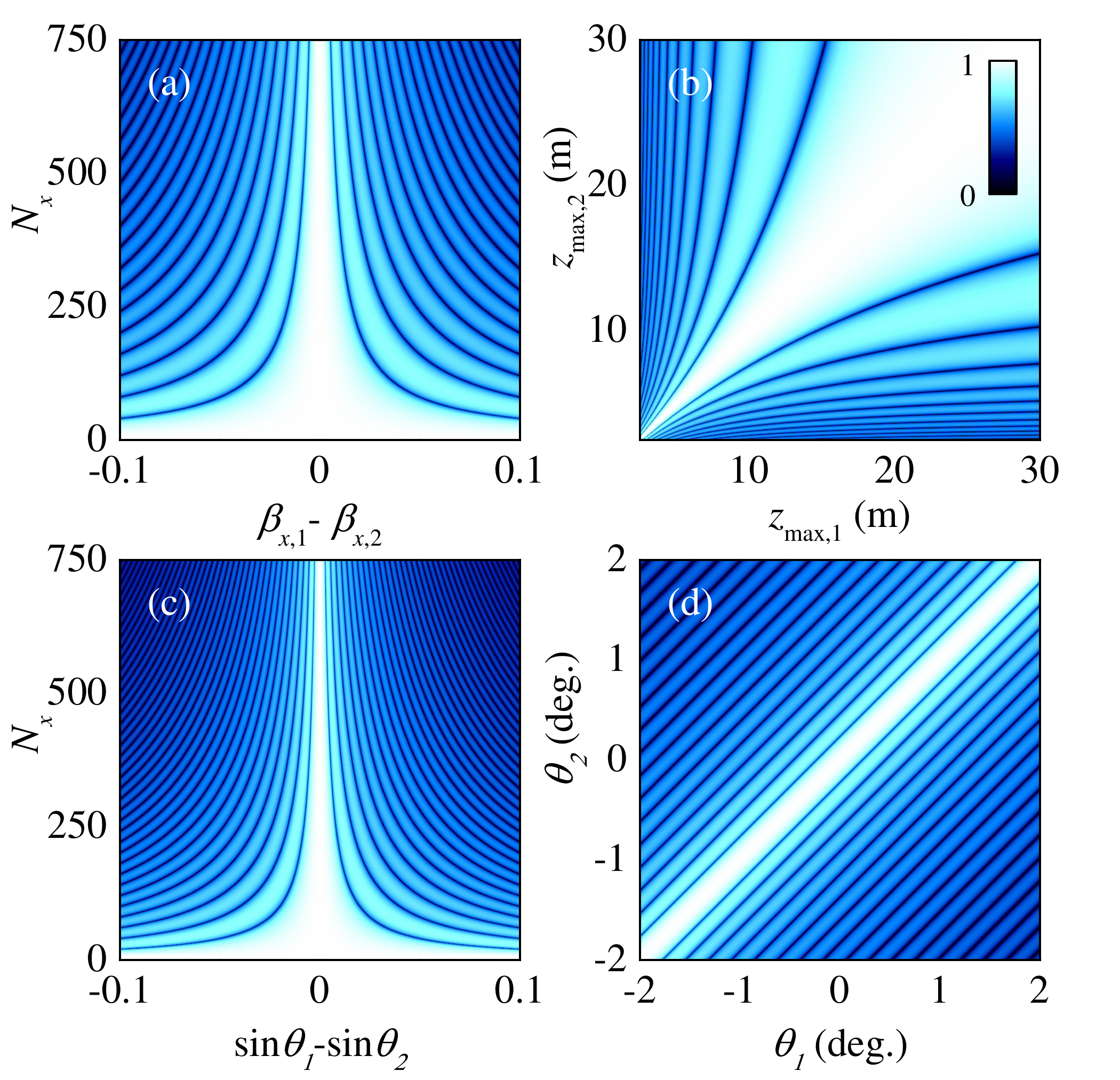}
    \caption{Correlation of cosine beams for $w_\theta=0$ or $w_z=0$. (a),(b) Beams with $w_\theta=0$, i.e., propagating along the same direction ($\theta_1=\theta_2$) and converging at different distance ($z_\mathrm{max,1}\neq z_\mathrm{max,2}$). (c),(d) Beams with $w_z=0$, i.e., propagating along different directions ($\theta_1\neq\theta_2$) and converging at the same distance ($z_\mathrm{max,1}=z_\mathrm{max,2}$). (a),(c) Global plots of $C_L$ for variable ULA size as a function of the detuning parameters. (b),(d) Examples for ULA with $N_x=500$ elements. The color plots are depicted in logarithmic scale, to emphasize the details.}
    	\label{fig:fig03}
\end{figure}

\noindent The case $w_\theta=0$ corresponds to zero detuning in the steering angle of the two beams. This is the case where two beams propagate along the same direction and converge at different distance $z_\mathrm{max}$, that is, beams with $\theta_1=\theta_2$ and $z_\mathrm{max,1}\neq z_\mathrm{max,2}$. For $w_\theta=0$, the summation in \eqref{Eq:EqCL} takes the simple form 
\begin{align}
C_L(0,w_z)= \left |\frac{1}{N_x}\sum_{n_x} e^{j w_z |n_x|} \right | = \left | \frac{\sin{\frac{N_x}{4}w_z}}{\frac{N_x}{2} \sin{\frac{1}{2}w_z}} \right|.
    \label{Eq:EqCLc}
\end{align}
The solutions of equation $C_L(0,w_z)=0$ satisfy 
\begin{align}
    w_z= \frac{4\pi p}{N_x} ,
    \label{Eq:EqCLcROOTSa}
\end{align}
where $p \in \mathbb{Z}$, with $p \neq 0, \pm N_x/2, \pm 2N_x/2,  \dots$, to exclude the nulls of the denominator that lead to $C_L(0,w_z)=1$. While $w_z$ can mathematically acquire any value, for practical applications, where $z_\mathrm{max}>N_xd_x/2$, it turns out that $w_z<kd_x$. According to \eqref{Eq:EqCLcROOTSa}, this range dictates that $|p|<kd_xN_x/4\pi$, which for typical elements with $\lambda/2$ spacing translates into $|p|<N_x/4$. Consequently, from the solution set \eqref{Eq:EqCLcROOTSa} only $p=0$ lifts the orthogonality, and we can formally write $p \in \mathbb{Z}_{\neq 0}$. \\
\indent In Fig.\ref{fig:fig03}(a) $C_L(0,w_z)$ is plotted as a function of the array size $N_x$ and the normalized detuning parameter $w_z/kd_x=\beta_{x,1}-\beta_{x,2}$, showing the first few branches within the range $p=\pm1,\pm2,\dots,\pm18$. Note how the number of orthogonal beam pairs increases with $N_x$. To investigate the nulls of $C_L(0,w_z)$ for a certain array size it is instructive to express the solution \eqref{Eq:EqCLcROOTSa} explicitly in terms of the individual convergence distances $z_\mathrm{max,1},z_\mathrm{max,2}$. Using \eqref{Eq:EqPAR1Db} we find
\begin{align}
    z_\mathrm{max,2} = \frac{N_x z_\mathrm{max,1}}{N_x-\frac{8\pi}{k d}\frac{z_\mathrm{max,1}}{N_x d}p},
    \label{Eq:EqCLcROOTSb}
\end{align}
where $p=\pm 1, \pm 2,\dots$, and in Fig.\,\ref{fig:fig03}(b) we plot $C_L(0,w_z)$ against $z_\mathrm{max,1},z_\mathrm{max,2}$ for an array of size $N_x=500$. \\
\indent While the result \eqref{Eq:EqCLcROOTSb} ensures the orthogonality of a single pair of beams, what is of interest is whether it also ensures general orthogonality among any pair. To investigate this possibility, we use beam 1 as reference beam with fixed properties, i.e. $z_\mathrm{max,ref}\equiv z_\mathrm{max,1}$. Then, all beams in \eqref{Eq:EqCLcROOTSb} are each one orthogonal to the reference beam and the set of $p's$ define the beamset with convergence distances given by \eqref{Eq:EqCLcROOTSb}. If we choose two random beams A and B from the beamset, they will both be orthogonal to the reference beam, satisfying 
\begin{subequations}
    \begin{align}
        &kd\left(\frac{N_x d}{2z_\mathrm{max,A}}-\frac{N_x d}{2z_\mathrm{max,ref}}\right) =  \frac{4\pi p_A}{N_x}, \label{Eq:EqPROOFa} \\
        &kd\left(\frac{N_x d}{2z_\mathrm{max,B}}-\frac{N_x d}{2z_\mathrm{max,ref}}\right) =  \frac{4\pi p_B}{N_x}, \label{Eq:EqPROOFb}
    \end{align}
\end{subequations}
where $p_A$ and $p_B$ are two different integers. Eliminating $z_\mathrm{max,ref}$ from \eqref{Eq:EqPROOFa},\eqref{Eq:EqPROOFb} leads to 
\begin{align}
    &kd\left(\frac{N_x d}{2z_\mathrm{max,A}}-\frac{N_x d}{2z_\mathrm{max,B}}\right) =  \frac{4\pi (p_A-p_B)}{N_x} \label{Eq:EqPROOFc},
\end{align}
where $p_A-p_B$ is also an integer and, therefore, beams A and B satisfy \eqref{Eq:EqCLcROOTSa}. Because beams A and B are different, $p_A-p_B$ is nonzero, ensuring that the denominator in \eqref{Eq:EqCLc} is nonzero. As a result, because the chosen beams can be any pair belonging to the beamset, this ensures that all beams are mutually orthogonal. \\
%
%
%
%
%
%
\subsubsection[Beams with zMAX1 = zMAX2]{Beams with $\theta_1\neq \theta_2$ and $z_\mathrm{max,1}=z_\mathrm{max,2}$}
\noindent The case $w_z=0$ corresponds to zero detuning in the convergence distance of the two beams. This is the case where two beams with the same $z_\mathrm{max}$ are directed towards different angles $\theta_1,\theta_2$, that is, beams with $\theta_1\neq \theta_2$ and $z_\mathrm{max,1}=z_\mathrm{max,2}$. For $w_z=0$, the summation in \eqref{Eq:EqCL} takes the simple form 
\begin{align}
C_L(w_\theta,0)= \left |\frac{1}{N_x}\sum_{n_x} e^{j w_\theta n_x} \right | = \left | \frac{\sin{\frac{N_x}{2}w_\theta}}{N_x \sin{\frac{1}{2}w_\theta}} \right|.
    \label{Eq:EqCLs}
\end{align}
The solutions of equation $C_L(w_\theta,0)=0$ satisfy 
\begin{align}
    w_\theta= \frac{2\pi q}{N_x} ,
    \label{Eq:EqCLsROOTSa}
\end{align}
where $q \in \mathbb{Z}$, with $q \neq 0, \pm N_x, \pm 2N_x,  \dots$, to exclude the nulls of the denominator that lead to $C_L(w_\theta,0)=1$. While $w_\theta$ can mathematically acquire any value, it corresponds to angular range and, therefore, it cannot exceed $2k d_x$ (see \eqref{Eq:EqPAR1Da}). Hence, the maximum achievable value for $w_\theta$ dictates that $|q|\leq N_x k d_x/\pi$, which for typical elements with $\lambda/2$ spacing translates into $|q|\leq N_x$. As a result, from the solutions set \eqref{Eq:EqCLsROOTSa} only $q=0$ lifts the orthogonality, hence $q \in \mathbb{Z}_{\neq 0}$. \\
\indent In Fig.\,\ref{fig:fig03}(c) $C_L(w_\theta,0)$ is plotted as a function of the array size $N_x$ and the normalized detuning parameter $w_\theta/kd_x=\sin{\theta_1}-\sin{\theta_2}$. Similarly to the case for $w_\theta=0$, there is an infinite countable set of $w_\theta$ values, for which the correlation is zero and the beams are orthogonal, for any array size $N_x$. To illustrate this property it is instructive to express the solution explicitly in terms of the individual angles $\theta_1,\theta_2$. Using \eqref{Eq:EqPAR1Da} in \eqref{Eq:EqCLsROOTSa} we find
\begin{align}
    \theta_2 = \arcsin \left ( \sin{\theta_1}-\frac{2\pi q}{N_x k d} \right ),
    \label{Eq:EqCLsROOTSb}
\end{align}
where $q=\pm 1, \pm 2,\dots$, and in Fig.\,\ref{fig:fig03}(d) we plot $C_L(w_\theta,0)$ against $\theta_1,\theta_2$ for an array of size $N_x=500$. Importantly, all beams belonging to the beamset \eqref{Eq:EqCLsROOTSb} are mutually orthogonal; the proof follows similar steps as in the case with $w_\theta=0$. \\
\subsubsection[Generally different beams]{Beams with $\theta_1\neq\theta_2$ and $z_\mathrm{max,1}\neq 
z_\mathrm{max,2}$}
Performing the summation in \eqref{Eq:EqCL} leads to the analytical result
\begin{align}
C_L(w_\theta,w_z)=\frac{1}{N_x}\frac{\sqrt{C_A+C_B+C_C}}{\left|\cos{w_z}-\cos{w_\theta}\right|},
    \label{Eq:EqCLformula}
\end{align}
where the terms $C_A.C_B,C_C$ are given explicitly by
\begin{subequations}
\begin{align}
    &C_A=\left(\cos{w_\theta}-\cos{w_z}\right)\left(\cos{(N_x w_\theta)}-1\right), \\
    &C_B=2\left(\cos{w_z}-1 \right)\left(\cos{w_\theta}+1 \right)\left(\cos{\frac{N_x w_z}{2}}\cos{\frac{N_x w_\theta}{2}}-1\right), \\
    &C_C=-2 \sin{w_z}\sin{w_\theta}\sin{\frac{N_x w_z}{2}}\sin{\frac{N_x w_\theta}{2}}. 
\end{align}
    \label{Eq:EqCLformulaCABC}  
\end{subequations}
\noindent For either $w_\theta=0$ or $w_z=0$, it is straightforward to verify that \eqref{Eq:EqCLformula} reduces to \eqref{Eq:EqCLc} and \eqref{Eq:EqCLs}, respectively. In the most general case, where $w_\theta\neq0,w_z\neq0$, the nulls of \eqref{Eq:EqCLformula}, can be found with simple inspection of the factorized form of terms $C_A,C_B,C_C$. Nulls of the numerator that occur with nulls of the denominator lead to nonzero correlation and, hence, it is required that the $|\cos{w_z}-\cos{w_\theta}|$ term remains nonzero when we are looking for the nulls of each individual term $C_A,C_B,C_C$ or of their sum $C_A+C_B+C_C$. With simple inspection of term $C_A$, the requirement for nonzero denominator and $C_A=0$ leads to $N_xw_\theta=2\pi q$, which is no other than the condition \eqref{Eq:EqCLsROOTSa}, previously found for the special case of $w_z=0$. The condition \eqref{Eq:EqCLsROOTSa}  leads also to $C_C = 0$ and, hence, \eqref{Eq:EqCLformula} reduces to
\begin{align}
    C_L=\frac{\sqrt{2\left(\cos{w_z}-1 \right)\left(\cos{ \frac{2\pi q}{N_x} }+1 \right)\left((-1)^q\cos{\frac{N_x w_z}{2}}-1\right)}}{N_x\left|\cos{w_z}-\cos{ \frac{2\pi q}{N_x} }\right|},
    \label{Eq:EqCLformulaB}
\end{align}
%
%
%
%
\begin{figure}[t!]
\centering
    \includegraphics[width=1\linewidth]{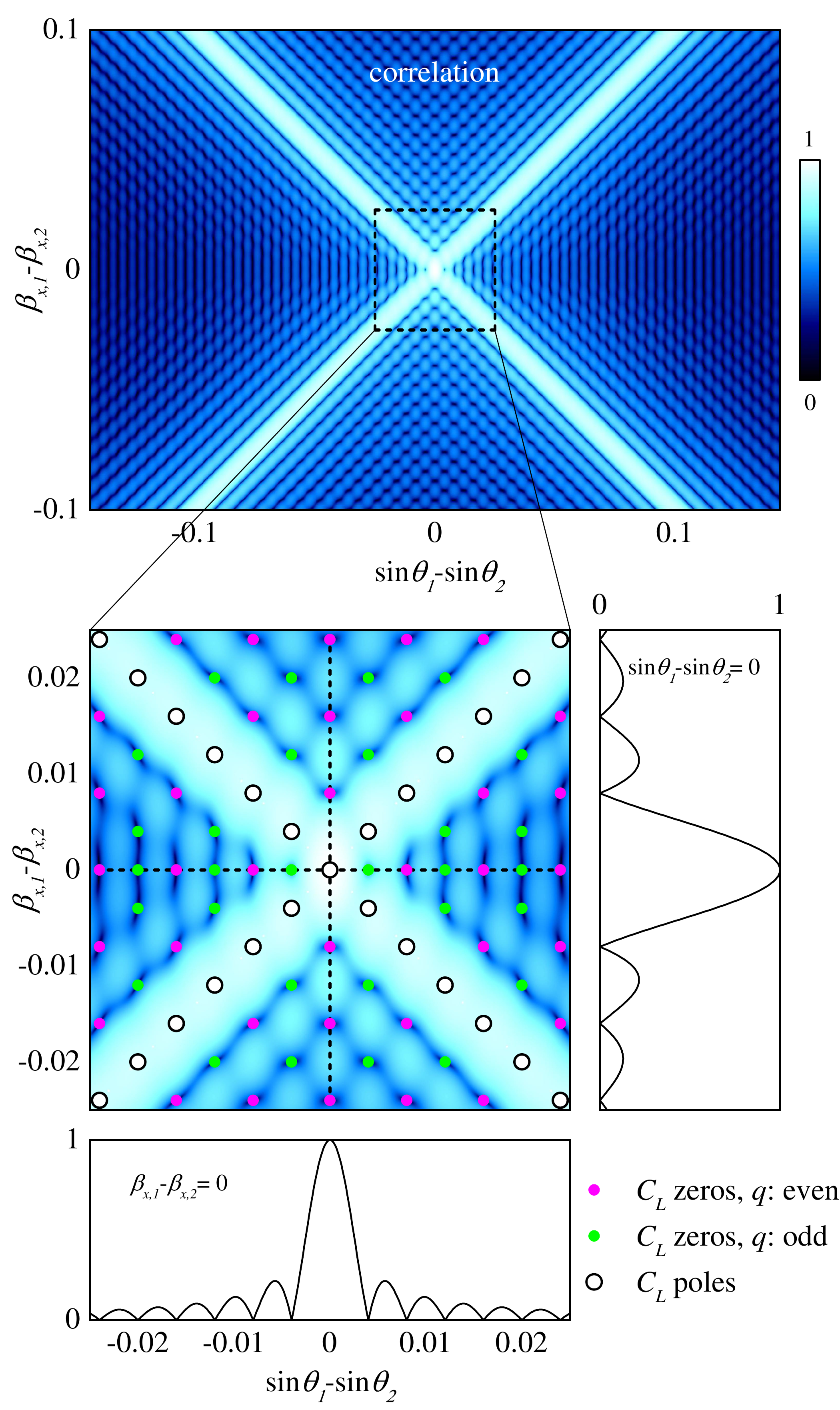}
	\caption{Correlation of cosine beams, as a function of the detuning parameters $w_\theta/kd_x= \sin{\theta_1} - \sin{\theta_2}$ and $w_z/kd_x=\beta_{x,1}-\beta_{x,2}$, for a ULA with $N_x=500$ elements. The zeros and poles of $C_L$ given by \eqref{Eq:EqCLROOTS} and \eqref{Eq:EqCLROOTSexclude}, respectively, are shown explicitly in a small region around $w_\theta=0,w_z=0$. The dashed lines mark the cross-sections at the characteristic detunings $w_\theta=0$ (right panel) and $w_z=0$ (bottom panel). The color plots are depicted in log scale, to emphasize the details.
 }   
    \label{fig:fig04}
\end{figure}
%
%
The nulls of the latter result can be found by individually setting to zero each of the three factors in the numerator of \eqref{Eq:EqCLROOTS}. Earlier we found that $w_z<kd_x$, which for typical $\lambda/2$ spacing translates into $w_z<\pi$. Hence, the first factor in the numerator is always nonzero, except for $w_z=0$. As a result, the nulls are primarily determined by the last factor, and are given by
  \begin{equation}
    w_z=
    \begin{cases}
      \frac{2\pi}{N_x}2p, & q:\text{even} \\
      \frac{2\pi}{N_x}(2p\pm 1), & q:\text{odd}
    \end{cases}
    \label{Eq:EqCLROOTS}
  \end{equation}
with
  \begin{equation}
    q\neq 
    \begin{cases}
      2 p, & q:\text{even},  \\
      2 p \pm 1, & q:\text{odd},
    \end{cases}
    \label{Eq:EqCLROOTSexclude}
  \end{equation}
where $p,q \in \mathbb{Z}$ and \eqref{Eq:EqCLROOTSexclude} excludes the nulls of the denominator that lead to nonzero $C_L$. Hence, under the constraint \eqref{Eq:EqCLROOTSexclude}, all beams with $w_\theta$ given by \eqref{Eq:EqCLsROOTSa} and $w_z$ given by \eqref{Eq:EqCLROOTS} guarantee that $C_L=0$. \\
\indent As an example, in Fig.\,\ref{fig:fig04} the correlation of two cosine beams with parameters ($\theta_1,z_\mathrm{max,1}$) and ($\theta_2,z_\mathrm{max,2}$), respectively, is shown for a ULA with $N_x=500$, as a function of the detunings $w_z/kd_x=\beta_{x,1}-\beta_{x,2}$ and $w_\theta/kd_x=\sin{\theta_1} - \sin{\theta_2}$. The zeros and poles of $C_L$ given by \eqref{Eq:EqCLROOTS} and \eqref{Eq:EqCLROOTSexclude}, respectively, are shown explicitly in a small region around $w_\theta=0,w_z=0$, together with cross-sections at the characteristic detunings $w_\theta=0$ (right panel) and $w_z=0$ (bottom panel).
\subsection{Beam correlations at the receiver}
\begin{figure}[t!]
\centering
    \includegraphics[width=1\linewidth]{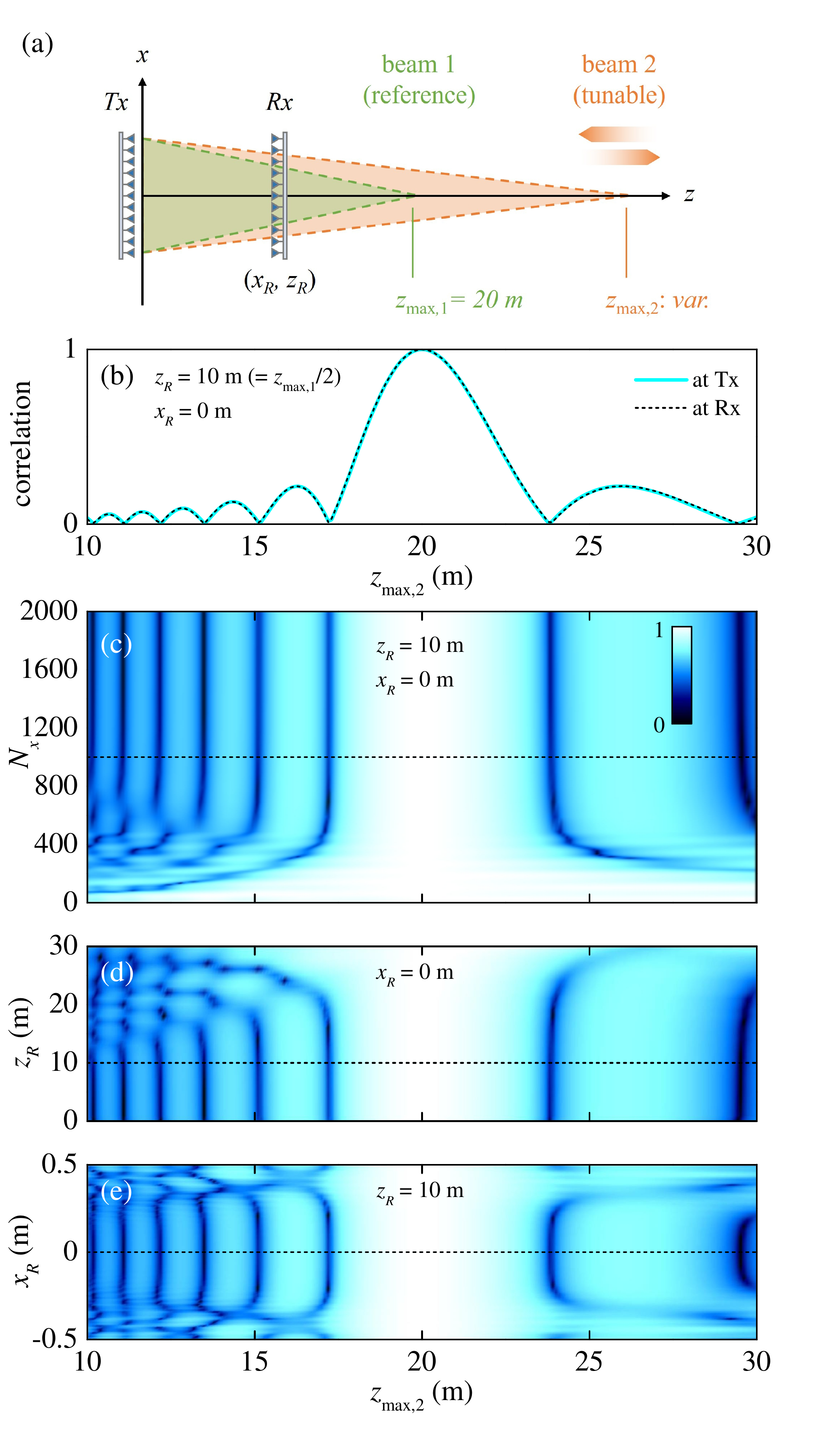}
	\caption{Beam orthogonality at the receiver. (a) Schematic illustration of two cosine beams generated by a Tx of size $N_x=1000$ and directed towards the Rx, located at ($x_R, z_R$). Beam 1 has fixed $z_\mathrm{max,1}=20\,\mathrm{m}$, and is used as reference. Beam 2 has tunable $z_\mathrm{max}$, accounting for a multitude of cosine beams. In this example $\theta_1=\theta_2=0^\circ$. (b) Beam correlation at the Tx and the Rx, as a function of $z_\mathrm{max,2}$, for Rx of size $N_x=1000$ located at ($x_R=0\,\mathrm{m},z_R=10\,\mathrm{m}$). (c) Beam correlation as a function of the Rx size. The dashed line marks the cross-section shown in (b). (d) Beam correlation as a function of the Rx displacement along $z$ ($x_R=0\,\mathrm{m}$) for Rx  size $N_x=1000$. (e) Beam correlation as a function of the Rx displacement along $x$ ($z_R=10\,\mathrm{m}$) for Rx size $N_x=1000$.}
      	\label{fig:fig05}
\end{figure}

\indent Having engineered the beam orthogonality at the Tx, in this section we verify that it is retrieved at the Rx. To this end, we propagate the beams numerically and we apply \eqref{Eq:EqCstep6} at the Rx.
\\
\indent Fist, we consider a Tx consisting of $N_x=1000$ elements and a Rx of the same size, located at $x_R=0,z_R=10\,\mathrm{m}$. For the reference beam we use $z_\mathrm{max,1}=20\,\mathrm{m}$ and we use beam 2 to scan $z_\mathrm{max,2}$, as shown schematically in Fig.\,\ref{fig:fig05}(a). In Fig.\,\ref{fig:fig05}(b) we plot \eqref{Eq:EqCLformula} (cyan solid line), i.e., the analytical correlation of the two beams at the Tx, and we also plot the numerically calculated correlation of the two beams at the Rx (blacked dashed line), as a function of $z_\mathrm{max,2}$. \\
\indent Next, in Fig.\,\ref{fig:fig05}(c) we scan the Rx size. As predicted by the derivation of \eqref{Eq:EqCstep6}, for small Rx panels where the beams are only partially captured by the Rx, the orthogonality is not retrieved correctly. However, with increasing Rx size, the received beams end up being entirely captured by the Rx, and the correlations converge to their analytical predictions. In fact, beyond a certain Rx size (here equal to the Tx size), the retrieved correlations do not change. \\
\indent Last, in Figs.\,\ref{fig:fig05}(d),(e) we test the retrieved orthogonality against longitudinal and lateral displacements, respectively, of the Rx. Due to the elongated shape of the beam, there is a large range of Rx positions along the $z$-axis, where the orthogonality is preserved; in turn, the correlations are more sensitive to lateral displacements and can be compensated with larger Rx panels. Importantly, our findings demonstrate the Rx can be located anywhere along the cosine beam range, which is deep in the near-field. Hence, by assigning different beams to different users, the concept of SDMA is brought to the near-field, hence the term NF-SDMA.
%
%
%
%
\section{Cosine beams with UPA's} \label{sec:CosineBeamsUPAs}
%
%
%
\begin{figure}[t!]
\centering
    \includegraphics[width=1\linewidth]{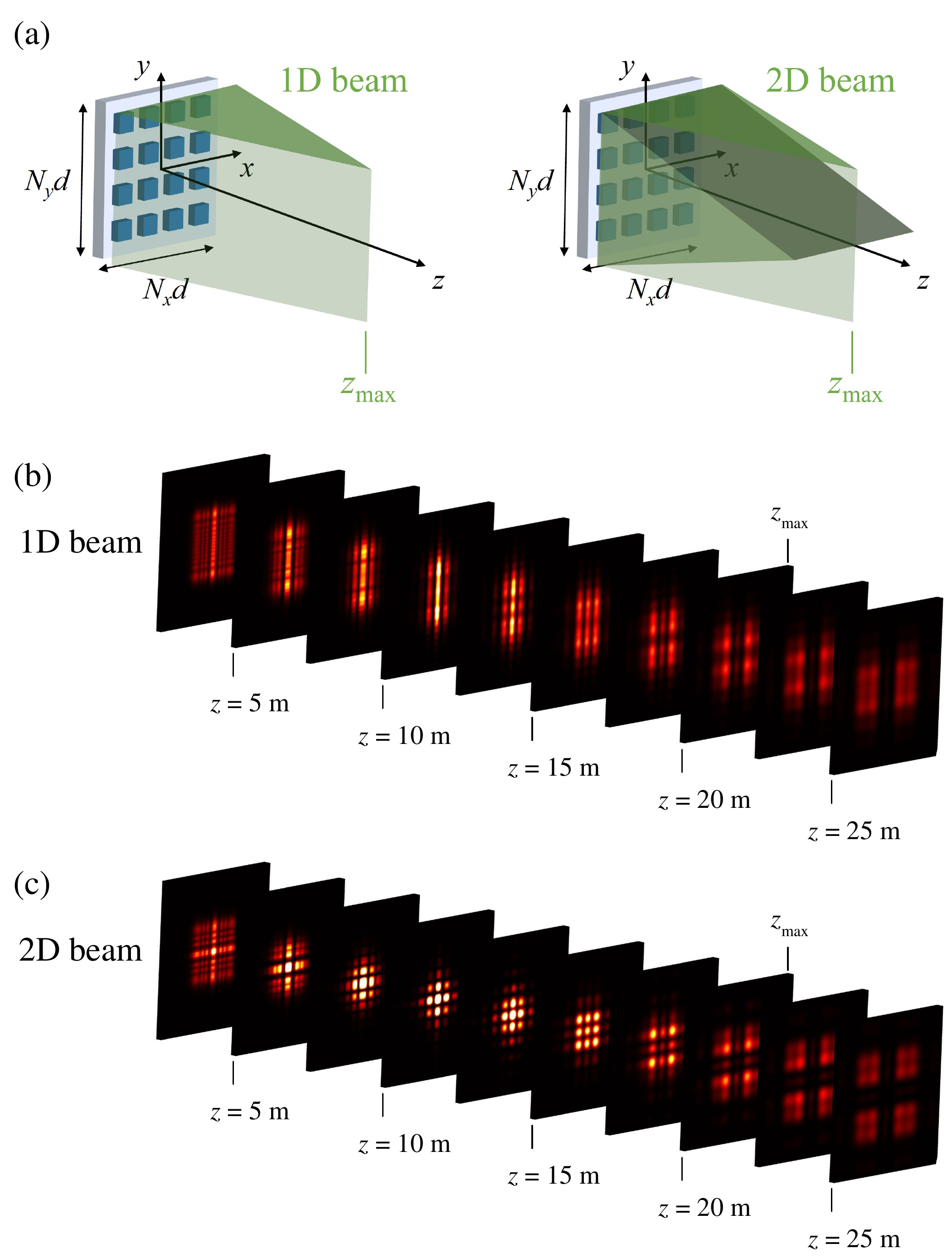}
	\caption{Generation of orthogonal cosine beams with UPA's. (a) Schematic representation of outer rays for the generation of 1D (left) and 2D (right) beams. Numerical propagation of (b) 1D and (c) 2D beam, with $z_\mathrm{max}=20\,\mathrm{m}$, using a UPA with $500\times500$ elements.}
      	\label{fig:fig06}
\end{figure}
%
%
\noindent In this section we extend the previous analysis from ULAs to UPAs (uniform planar arrays). The UPA is centered at the origin of the coordinate system, extending on the $xy$-plane, and consisting of $N_x \times N_y$ elements, periodically arranged with periodicity $d_x$ and $d_y$ along the $x$ and $y$ directions, respectively. The UPA generates beams with tailored amplitude and phase characteristics that propagate in the $z>0$ semi-infinite space. The $E$-field profile at the input plane ($z=0$) has the form
\begin{align}
    E(x,y,0)=A(x,y) e^{j\phi(x,y)},
    \label{Eq:EqINPUTWAVE2D}
\end{align}
For cosine beams, the phase profile \eqref{Eq:EqIMPUTphase1Dcont} is generalized to two dimensions as 
\begin{align}
    \phi(x,y)=k\sin{\theta}\cos{\phi}x+k\sin{\theta}\sin{\phi}y-k\beta_x|x|-k\beta_y|y|,
    \label{Eq:EqIMPUTphase2Dcont}
\end{align}
where the angles $\theta,\phi$ account for the steering direction and the parameters $\beta_x,\beta_y$ control the beam convergence on the $xz$ and $yz$ plane, respectively. The latter are expressed in terms of the convergence distance as
\begin{subequations}
    \begin{align}
        &\beta_x=\frac{N_x d_x}{2z_{\mathrm{max},x}} \label{Eq:EqBETAx} \\
        &\beta_y=\frac{N_y d_y}{2z_{\mathrm{max},y}} \label{Eq:EqBETAy},
    \end{align}
\end{subequations}
where $z_{\mathrm{max},x},z_{\mathrm{max},y}$ is the convergence distance on the $xz$ and $yz$ plane, respectively. Without loss of generality we will consider beams with $z_{\mathrm{max},x} = z_{\mathrm{max},y} \equiv z_\mathrm{max}$. \\
\indent To derive the condition for beam orthogonality, we discretize the continuous phase on the UPA grid, i.e. we express \eqref{Eq:EqIMPUTphase2Dcont} at points $x=n_xd_x,y=n_yd_y$. Then, the correlation of steering vectors for two beams characterized by ($\theta_1,\phi_1,z_\mathrm{max,1}$) and ($\theta_2,\phi_2,z_\mathrm{max,2}$), respectively, is formulated as
\begin{multline}
    C_P(w_{\theta x},w_{\theta y},w_{z x},w_{z y}) = \\
    \left |\frac{1}{N_xN_y}\sum_{n_x}\sum_{n_y} e^{j w_{\theta x} n_x}e^{j w_{\theta y} n_y}e^{j w_{z x} |n_x|}e^{j w_{z y} |n_y|} \right |,
    \label{Eq:EqCORR_UPA}
\end{multline}
where the subscript $P$ denotes that the correlation function refers to a planar array. The involved parameters are defined as
\begin{subequations}
    \begin{align}
        &w_{\theta x}=kd_x(\sin{\theta_1}\cos{\phi_1}-\sin{\theta_2}\cos{\phi_2}) \\
        &w_{\theta y}=kd_y(\sin{\theta_1}\sin{\phi_1}-\sin{\theta_2}\sin{\phi_2}) \\
        &w_{z x}=kd_x(\beta_{x,1}-\beta_{x,2})=kd_x\left(\frac{N_x d_x}{2z_\mathrm{max,1}}-\frac{N_x d_x}{2z_\mathrm{max,2}}\right) \\
        &w_{z y}=kd_y(\beta_{y,1}-\beta_{y,2})=kd_y\left(\frac{N_y d_y}{2z_\mathrm{max,1}}-\frac{N_y d_y}{2z_\mathrm{max,2}}\right),
    \end{align}
    \label{Eq:EqPAR2D}
\end{subequations}
and
\begin{subequations}
    \begin{align}
        &n_x = -\frac{N_x-1}{2},-\frac{N_x-1}{2}+1,\dots,\frac{N_x-1}{2} \\
        &n_y = -\frac{N_y-1}{2},-\frac{N_y-1}{2}+1,\dots,\frac{N_y-1}{2} \\
    \end{align}
    \label{Eq:EqNxNy}
\end{subequations}
Because the contributions of the $x$ and $y$ components to the sum \eqref{Eq:EqCORR_UPA} are independent, we may group the summation as
\begin{multline}
        C_P(w_{\theta x},w_{\theta y},w_{z x},w_{z y}) = \\
        \left |\left(\frac{1}{N_x}\sum_{n_x}e^{j w_{\theta x} n_x} e^{j w_{z x} |n_x|} \right) \left( \frac{1}{N_y}\sum_{n_y}e^{j w_{\theta y} n_y} e^{j w_{z y} |n_y|} \right) \right |,
\end{multline}
which is simply written in terms of $C_L$, the correlation function for ULA's, as
\begin{align}
    C_P(w_{\theta x},w_{\theta y},w_{z x},w_{z y}) = C_L(w_{\theta x},w_{z x})C_L(w_{\theta y},w_{z y}).
    \label{Eq:EqCPformula}
\end{align}
For steering on the $xz$-plane we set $\phi = 0$, which leads to $w_{\theta x}=w_\theta$ and $w_{\theta y}=0$. \\ 
\indent For $w_{z y}=0$ we can shape beams that converge along the $x$-direction and are invariant along the $y$-direction, i.e. 1D cosine beams, using the entire two-dimensional (2D) surface of the UPA, as we schematically show in Fig.\,\ref{fig:fig06}(a), left panel. In Fig.\,\ref{fig:fig06}(b) we demonstrate the numerical propagation of a 1D beam with $\theta=0$ and $z_\mathrm{max}=20\,\mathrm{m}$. Taking into account that $C_L(0,0)=1$ it is straightforward to see that the correlation function $C_P$ of 1D cosine beams reduces to that of a ULA. Hence, with UPAs, all the conclusions previously drawn about beam orthogonality in ULA's hold here as well. Importantly, due to the 2D nature of UPAs, the beam divergence in the $y$-direction is significantly reduced with respect to ULAs, as verified in the snapshots shown in Fig.\,\ref{fig:fig06}(b). Hence, with UPAs the beam can be concentrated within a narrow aperture, maximizing the power at the Rx that resides in its near-field (the Fraunhofer distance in this example is $250\,\mathrm{m}$). \\
\indent For $w_{zy}\neq0$ we take full advantage of the 2D nature of UPAs to form 2D cosine beams, i.e. beams that converge on both the $xz$ and $yz$ planes, as we schematically show in Fig.\,\ref{fig:fig06}(a), right panel. In Fig.\,\ref{fig:fig06}(c) we demonstrate the numerical propagation of such a beam that bears all the properties of 1D cosine beams and also has the advantage of independently controlling the beam orthogonality in the two transverse planes.
\section{Codebook design for NF-SDMA} \label{sec:CodebookA}
\noindent To design codebooks for NF-SDMA, we need to construct a set of mutually orthogonal beans, i.e. to define beams with properties that correspond to the nulls of \eqref{Eq:EqCLformula} for any pair combination. Previously we showed that $w_\theta$ given by \eqref{Eq:EqCLsROOTSa} and $w_z$ by \eqref{Eq:EqCLROOTS} guarantee that $C_L=0$, with the exception of solutions \eqref{Eq:EqCLROOTSexclude}. 
For this beamset we have already demonstrated that beams that either (a) have the same $z_\mathrm{max}$ along any $\theta$ ($w_z = 0$) or (b) have the same $\theta$ with any $z_\mathrm{max}$ ($w_\theta = 0$), are all mutually orthogonal; this beamset also has the advantage that the angles $\theta$ can be selected independently of the distances $z_\mathrm{max}$. 
%
Next, we will show that, in the general case where $w_\theta \neq 0, w_z \neq 0$, the pair-wise orthogonality that characterizes this beamset can be satisfied for many combinations of such beams, enabling a plethora of communication modes. \\
\indent First, using \eqref{Eq:EqCLsROOTSa}, we will enumerate all the possible beam directions within the angular range of interest. Without loss of generality we may set $\theta_2=0$ for the reference beam, and identify all beam directions $\theta\equiv\theta_1$ contained within the angular range defined by the maximum angle $\theta_\mathrm{max}$, i.e. count all beams that satisfy $\theta\leq\theta_\mathrm{max}$.
The maximum integer, $q_\mathrm{max} \equiv \mathrm{max}(q)$, that satisfies this condition is found by using \eqref{Eq:EqPAR1Da} to solve \eqref{Eq:EqCLsROOTSa} in terms of $q$, and is given by
\begin{align}
     q_\mathrm{max} =\lfloor\sin{(\theta_\mathrm{max})}\frac{N_x k d_x}{2\pi}\rfloor,
     \label{Eq:EqQmax}
\end{align}
%
where ${\lfloor \cdot \rfloor}$ is the floor function. Hence, the set of orthogonal beam directions is given by the angles
\begin{align}
    \theta =  \arcsin{\left( \frac{2 \pi q}{N_x kd_x} \right)}
    \label{Eq:EqCLsROOTSc}
\end{align}
with
\begin{align}
    q =  -q_\mathrm{max}, \dots, -1,0,1, \dots,  q_\mathrm{max}.
\end{align}
%
%
%
\begin{figure}[t!]
\centering
    \includegraphics[width=1\linewidth]{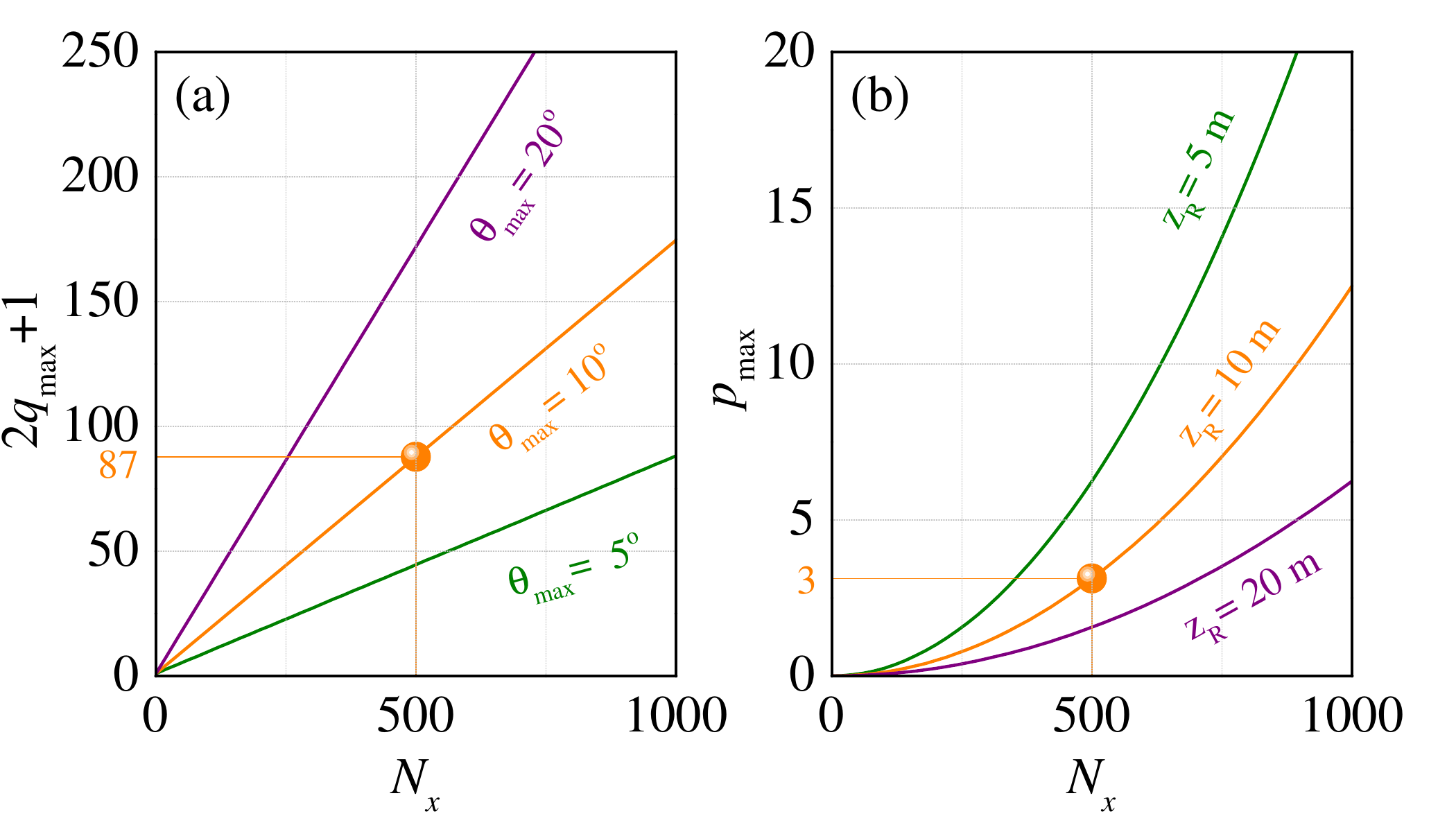}
	\caption{Maximum number of communication modes, as a function of the array size. (a) Modes per maximum steering angle and examples for $\theta_\mathrm{max}=5^\circ,10^\circ,20^\circ$. (b) Modes per minimum convergence distance and examples for $z_\mathrm{R}=5\,\mathrm{m},10\,\mathrm{m},20\,\mathrm{m}$. The filled dot marks the case for $N_x=500$ and $\theta_\mathrm{max}=10^\circ$, $z_\mathrm{R}=10\,\mathrm{m}$.}
    \label{fig:fig07}
\end{figure}
%
%
\indent At this point, it is important to note that the steering angles in \eqref{Eq:EqCLsROOTSc} are also the uncorrelated steering angles in typical SDMA. This is a direct consequence of the fact that cosine beams are formed by a pair of counter-propagating waves; in essence, they are linear superposition of the beams used in SDMA and, as such, they bear the same correlation properties. \\
\indent Next, we will enumerate the beams along each direction chosen in the previous step, using \eqref{Eq:EqCLROOTS}. Taking into account that the range of any beam must be larger than $z_\mathrm{R}$, the Rx distance from the Tx, for beams 1 and 2 it must hold $z_\mathrm{max_1},z_\mathrm{max_2}\geq z_\mathrm{R}$. Using beam 2 as reference beam and setting $ z_\mathrm{max,2}\rightarrow \infty$, ensures that the reference beam range always satisfies the above condition. Hence, we search for all beam ranges $z_\mathrm{max}\equiv z_\mathrm{max,1}$ to identify the beams that satisfy $z_\mathrm{max}\geq z_\mathrm{R}$.
The maximum integer, $p_\mathrm{max} \equiv \mathrm{max}(p)$, that satisfies this condition is found by using \eqref{Eq:EqPAR1Db} to solve \eqref{Eq:EqCLROOTS} in terms of $p$, and is given by
  \begin{equation}
    p_\mathrm{max} 
    \begin{cases}
      \lfloor \frac{kd_x}{\pi} \frac{d_x N_x^2}{8 z_\mathrm{R}}\rfloor, & q:\text{even} \\
      \lfloor \frac{kd_x}{\pi} \frac{d_x N_x^2}{8 z_\mathrm{R}}+\frac{1}{2}\rfloor, & q:\text{odd}
    \end{cases}
    \label{Eq:EqPmax}
  \end{equation}
Note that, in \eqref{Eq:EqCLROOTS}, both signs in the lower branch lead to the same solution set with enumeration starting from $p=0$ for the $"-"$ sign and from $p=1$ for the $"+"$ sign. In \eqref{Eq:EqPmax} we chose the $"+"$ sign to have common enumeration in both branches. Hence, the set of orthogonal beam ranges is given by the distances
  \begin{equation}
    z_\mathrm{max}=
    \begin{cases}
      \frac{kd_x}{\pi} \frac{d_x N_x^2}{8p}, & q:\text{even} \\
      \frac{kd_x}{\pi} \frac{d_x N_x^2}{4 (2p - 1)}, & q:\text{odd}
    \end{cases}
    \label{Eq:EqCLcROOTSc}
  \end{equation}
with
\begin{align}
    p =  1, 2, \dots,  p_\mathrm{max}.
\end{align}
As a result, the maximum number of communication modes, $M_\mathrm{max}$, for a certain choice of $q_\mathrm{max},p_\mathrm{max}$ is 
\begin{align}
    M_\mathrm{max}=(2q_\mathrm{max}+1)p_\mathrm{max}
\end{align}
In Fig.\,\ref{fig:fig07} we demonstrate the maximum number of possible communication modes per (a) maximum steering angle and (b) minimum convergence distance, as a function of the array size. Note that $q_\mathrm{max}$ scales linearly with $N_x$, while $p_\mathrm{max}$ is quadratic in $N_x$.\\
\indent All beams defined by directions \eqref{Eq:EqCLsROOTSc} and ranges \eqref{Eq:EqCLcROOTSc} are orthogonal with respect to the reference beam ($\theta_2=0$,$z_\mathrm{max,2}\rightarrow\infty$) and, therefore, \eqref{Eq:EqCLsROOTSc}, \eqref{Eq:EqCLcROOTSc} are necessary, though not sufficient conditions for mutual beam orthogonality. We can now select a random pair of two such beams to examine their mutual orthogonality. From \eqref{Eq:EqCLsROOTSc} we choose two random angles, characterized by the integers $q_A$ and $q_B$, corresponding to beams A and B, respectively. Similarly, from \eqref{Eq:EqCLcROOTSc} we choose two random ranges, characterized by the integers $p_A$ and $p_B$, corresponding to beams A and B, respectively. In terms of the integers $q,p$ that enumerate the angles and ranges, beam A is characterized by $q_A,p_A$ and beam B by $q_B,p_B$. Inserting the chosen parameters in \eqref{Eq:EqPAR1Da} and \eqref{Eq:EqPAR1Db}, and inserting the parameters $w_\theta,w_z$ in \eqref{Eq:EqCLformula} leads to the following conditions for mutual beam orthogonality
%
%
%
\begin{figure}[t!]
\centering
    \includegraphics[width=1\linewidth]{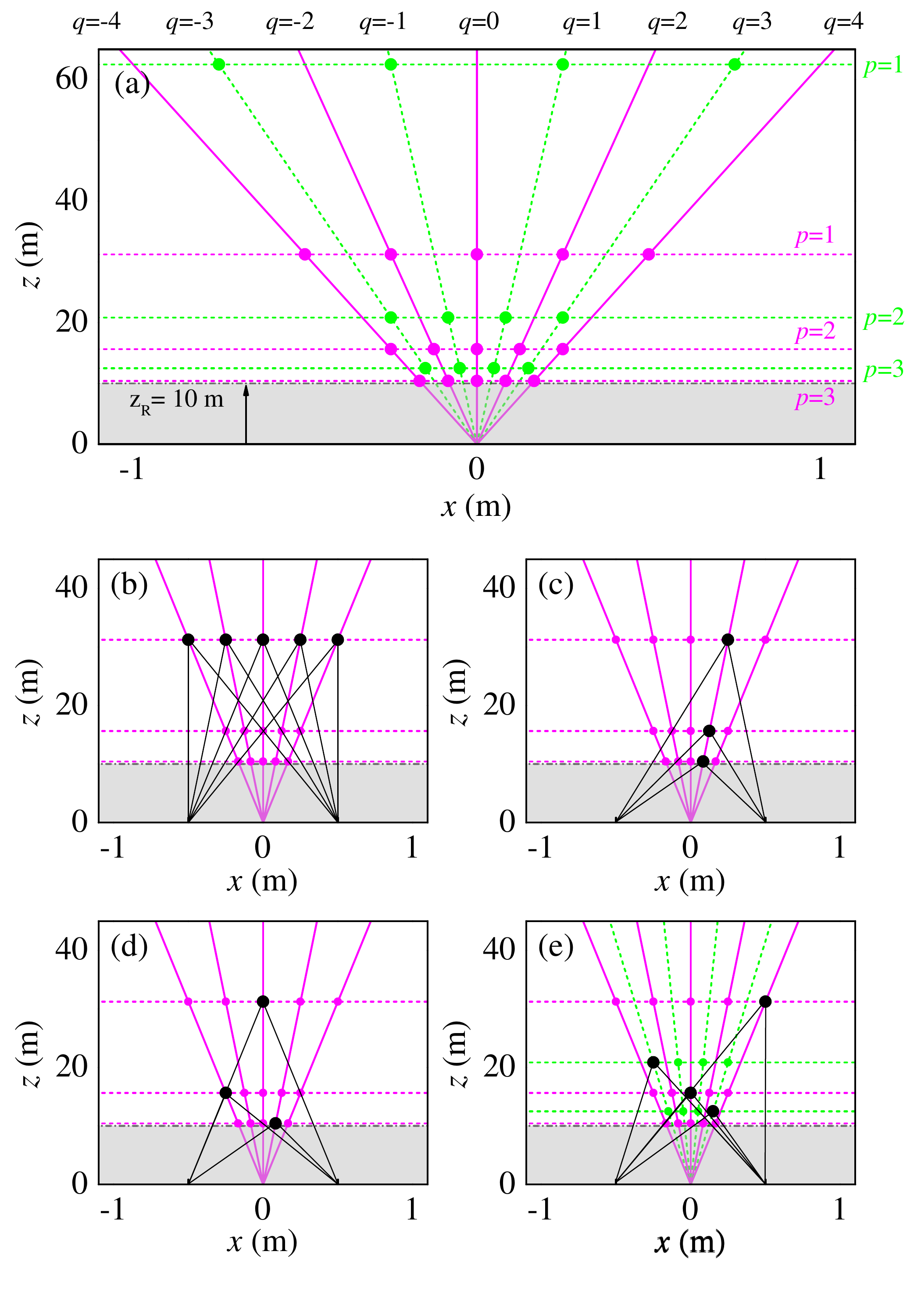}
	\caption{Codebook design and examples of communication modes for Tx size of $N_x=500$. (a) The convergence distance for each $p$ is marked with the horizontal dashed lines ($p_\mathrm{max}=3$) (magenta for $q$:even, green for $q$:odd). The steering direction for each $q$ is marked with the solid lines, showing the first 9 out of total of $2 q_\mathrm{max}+1=87$ steering directions. The shaded area marks the region where the Rx is located. (b),(c),(d),(e) Examples of sets of communication modes, marked with the black dots. Each triangle marks the corresponding cosine beam range of each mode. (b) Example set for $w_z=0$. (c) Example set for $w_\theta=0$. (d) Example set with modes on $q$:even. (e) Example set with modes on $q$ of mixed parity.} 
    \label{fig:fig08}
\end{figure}
%
%
  \begin{equation}
    w_z=
    \begin{cases}
      \frac{4\pi}{N_x}(p_A-p_B), & q_A-q_B:\text{even}, \\
      \frac{4\pi}{N_x}(p_A-p_B+\frac{1}{2}), & q_A-q_B:\text{odd},
    \end{cases}
    \label{Eq:EqMUTUAL}
  \end{equation}
with
  \begin{equation}
    |q_A-q_B|\neq 
    \begin{cases}
      2|p_A-p_B|, & q_A-q_B:\text{even},  \\
      2|p_A-p_B+\frac{1}{2}|, & q_A-q_B:\text{odd}.
    \end{cases}
    \label{Eq:EqMUTUALexclude}
  \end{equation}
%
%
%
Because the difference of any two even or odd integers is always an even integer, the condition $"q_A-q_B:\text{even}"$ refers to any beams on directions defined by \eqref{Eq:EqCLsROOTSc} with either even or odd $q$. Similarly, the condition $"q_A-q_B:\text{odd}"$ refers to any pair of beams, with $q$ of mixed parity. \\
%
%
%
\indent As an example, let us define the communication modes within a narrow angular range of $10^\circ$, for users residing up to $z_\mathrm{R}=10\,\mathrm{m}$ far from the Tx, which has size $N_x=500$. Setting $\theta_\mathrm{max}=10^\circ$ and $z_\mathrm{R}=10\,\mathrm{m}$, leads to $q_\mathrm{max}=87$ and $p_\mathrm{max}=3$, also marked in Fig.\,\ref{fig:fig07} with the filled dots. Hence, there is a maximum of $M_\mathrm{max}=261$ possible communication modes. A codebook design using only the first few modes is shown in Fig.\,\ref{fig:fig08}. Fig.\,\ref{fig:fig08}(a) depicts the spatial distribution of all possible communication modes within a small angular range of $\pm0.92^\circ$, corresponding to the steering directions $q=-4,\dots,+4$ and all convergence distances, i.e. $p=1,2,3$. The magenta (green) lines denote that $q$ is even (odd). Each dot denotes one such mode, which is associated with a single ($q,p$) pair. The location of each dot also marks the cosine beam range of the beam that corresponds to the specific mode. The shaded area marks the region where the Rx is located. Examples of sets of communication modes are depicted in Figs.\,\ref{fig:fig08}(b)-(e), marked with the black dots, also showing the corresponding cosine beam range of each mode, which is contained in the triangular region of each mode. In Fig.\,\ref{fig:fig08}(b) five modes with the same $z_\mathrm{max}$ are depicted ($w_z=0$), characterized by $p=1$ and $q=-4,-2,0,2,4$. In Fig.\,\ref{fig:fig08}(c) three modes with the same $\theta$ are depicted ($w_\theta=0$), characterized by $p=1,2,3$ and $q=2$. In Fig.\,\ref{fig:fig08}(d) the three modes all reside on steering angles with $q$:even, and are characterized by ($q=-4,p=2$), ($q=0,p=1$) and ($q=2,p=3$). Last, in Fig.\,\ref{fig:fig08}(e) the four modes reside on steering angles of mixed parity, and are characterized by ($q=-3,p=2$), ($q=0,p=2$), ($q=3,p=3$) and ($q=4,p=1$).\\

\section{Codebook design for single antenna receivers} \label{sec:CodebookB}
%
%
%
\begin{figure}[t!]
\centering
    \includegraphics[width=1\linewidth]{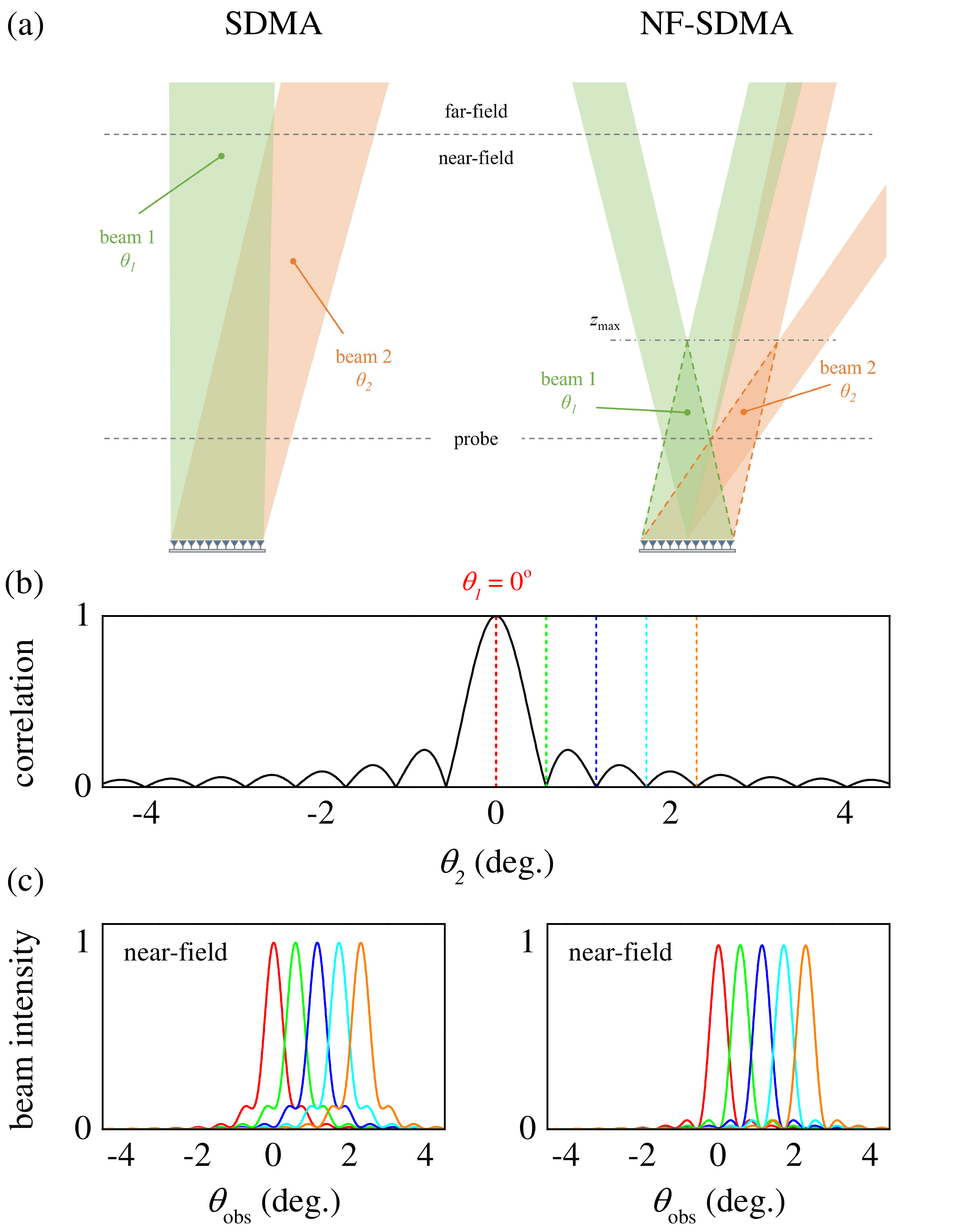}
	\caption{Near-field beam orthogonality for single antenna receivers. (a) Schematic illustration of two uncorrelated beams, in typical SDMA (left) and in NF-SDMA (right). (b) Correlation of beam 1 with beam 2, shown in (a), for different steering angles. The reference beam 1 is directed towards $\theta_1=0^\circ$. For NF-SDMA the reference beam 1 converges at $z_\mathrm{max}=20\,\mathrm{m}$ (c) Near-field beam at $z=10\,\mathrm{m}$. The ULA consists of 200 elements (with $\lambda/2$ spacing), leading to a Fraunhofer distance of $40\,\mathrm{m}$.}    
    \label{fig:fig09}
\end{figure}
%
%
\noindent Typically, because the propagation of beams in the near-field is accompanied by intensity oscillations that change qualitatively with the propagation distance, it is difficult to have an SDMA analogue in the near-field. Note, however, that because cosine beams are characterized by a standing wave pattern that extends across the entire cosine beam range, their near-field is predictable and can be tailored to promote or suppress the beam intensity at selected regions. Hence, it is possible to design codebooks, where the intensity at the maximum of one beam overlaps with zeros from all the other beams of the codebook. This possibility essentially brings the typical SDMA in the near-field, which is an ideal scheme for receivers equipped with single antennas, instead of arrays. \\
%
%
\indent The standing wave that is formed in a cosine beam with parameter $\beta_x$ has the form $\cos^2{\left(k \beta_x x\right)}$ and extends along the cosine beam range. Mathematically, a perfect standing wave requires two counter-propagating plane waves of infinite extent. However, due to the finite size of the Tx, the cosine wave is formed by a truncated version of those waves, and the finite aperture of the Tx introduces an envelope that suppresses the peaks of the standing wave towards the edges \cite{Droulias2018}. Depending on the ratio of $z_\mathrm{max}$ relative to the size of the array, which is expressed via $\beta_x$, the envelope can controllably shrink, in turn suppressing the peaks of the standing wave. In fact, at $z=z_\mathrm{max}/2$, the two waves interfere across the entire beam cross-section, forming a standing wave of $N_xd_x/2$ extent, beyond which the beam intensity quickly decays to zero (see Fig.\,\ref{fig:fig02}). 
%
The first zeros of $\cos^2{\left(k \beta_x x\right)}$ occur at $x=\pm\pi/(2k\beta_x)$. To suppress interference with beams at other angles we tune the first zeros of the standing wave to occur at $x=\pm N_xd_x/2$, leading to 
\begin{align}
    z_\mathrm{max}=\frac{(N_xd_x)^2}{\lambda}\equiv \frac{z_F}{2},
\end{align}
where $z_F$ is the Fraunhofer distance. Probing the beam at $z=z_\mathrm{max}/2$ translates to $z=z_F/4$, which resides in the radiating near-field. At such short distance,  we will demonstrate that NF-SDMA provides interference suppression; on the contrary, typical SDMA fails, because SDMA beams have not yet reached their far-field form. \\
\indent In Fig.\,\ref{fig:fig09} we compare typical SDMA with our proposed NF-SDMA scheme for a Tx of size $N_x=200$, for which $z_F=40\,\mathrm{m}$. Following the preceding analysis we use $z_\mathrm{max}=z_F/2=20\,\mathrm{m}$ and we probe all beams at $z=z_F/4=10\,\mathrm{m}$. A schematic representation of both schemes is depicted in Fig.\,\ref{fig:fig09}(a), where the colored zones represent rays. The beam correlations for both schemes are calculated using \eqref{Eq:EqCLformula} with $w_\theta$ given by \eqref{Eq:EqCLs} and $w_z=0$; they are presented as a function of the steering angle in Fig.\,\ref{fig:fig09}(b), where the reference beam has $\theta=0$. The dashed colored lines denote the steering angles of the beams used under both schemes. In Fig.\,\ref{fig:fig09}(c), cross-sections of the beams under both schemes are shown, where the color-code follows that in panel (b). Because all beams are probed deep in the near-field, the local orthogonality in SDMA is lost. On the contrary, in NF-SDMA, all beams are locally orthogonal, essentially bringing the benefits of typical SDMA into the near-field. \\
\section{Conclusion} \label{sec:Conclusion}
\noindent In this work, we introduced the concept of NF-SDMA, to realize multiple access in the near-field, for applications in future wireless connectivity. By judicious design, we selected the family of cosine beams as an ideal beamset for realizing NF-SDMA, and we demonstrated that such beams offer the necessary orthogonality (strictly zero correlations), from the near- up to the far-field of the transmitter. We derived analytically their correlations and demonstrated analytically and verified numerically the plethora of communication modes offered by these beams, for both ULAs and UPAs. Based on our findings, we proposed codebook designs for NF-SDMA, for both multi-antenna and single-antenna receivers. With our work, we aim to lay the ground for multiple access in the near-field, for 6G wireless connectivity and beyond.
\section*{Acknowledgments}
This work was supported by the European Commission’s Horizon Europe Programme under the Smart Networks and Services Joint Undertaking INSTINCT project (Grant Agreement $101139161$).
%
%
\appendix[ ]
\subsection[Beam propagation in k-space]{Beam propagation in $k$-space}
\label{Sec:AppendixA}
\noindent Let us denote as $\textbf{\textrm E}(\textbf{\textrm r}) = \textbf{\textrm E}(x, y, z)$ the $E$-field of a beam at the observation point $\textbf{\textrm r}$=($x,y,z$). The beam propagates along $z$ and, at $z=0$ where it is generated by the Tx, its $E$-field is $\textbf{\textrm E}(x, y, 0)\equiv \textbf{\textrm E}_\mathrm{T}(x, y)$. Its two-dimensional Fourier transform is

\begin{align}
    \hat{\textbf{\textrm E}}_\mathrm{T}(k_x,k_y) = \frac{1}{4\pi^2} \iint_{-\infty}^{+\infty} \textbf{\textrm E}_\mathrm{T}(x,y) e^{-j(k_xx+k_yy)}dxdy,
    \label{Eq:EqA1}
\end{align}

where $x, y$ are the Cartesian transverse coordinates and $k_x, k_y$ the corresponding spatial frequencies. Similarly, the inverse Fourier transform reads

\begin{align}
    \textbf{\textrm E}_\mathrm{T}(x,y) = \iint_{-\infty}^{+\infty} \hat{\textbf{\textrm E}}_\mathrm{T}(k_x,k_y)  e^{j(k_xx+k_yy)}dk_xdk_y.
    \label{Eq:EqA2}
\end{align}

Note that the field $\textbf{\textrm E}_\mathrm{T}$ and its Fourier transform $\hat{\textbf{\textrm E}}_\mathrm{T}$ represent vectors and, hence, the Fourier integrals hold separately for each vector component. The field has to satisfy Maxwell's equations, which for free-space propagation reduce to the vector Helmholtz equation ($ \nabla^2+k^2) \textbf{\textrm E}(\textbf{\textrm r})=0$. Expressing similarly the field $\textbf{\textrm E}(x, y, z)$ via its Fourier transform and inserting it into the Helmholtz equation, we find that the Fourier spectrum $\hat{\textbf{\textrm E}}_\mathrm{R}$ of the beam at distance $L$, where the Rx is located, is given by

\begin{align}
    \hat{\textbf{\textrm E}}_\mathrm{R}(k_x,k_y;L) = \hat{\textbf{\textrm E}}_\mathrm{T}(k_x,k_y) e^{jk_zL},
    \label{Eq:EqA3}
\end{align}

where $k_z=\sqrt{k^2-k_x^2-k_y^2}$ and $k$ is the wavenumber. \\
\subsection{Input phase for the generation of cosine beams}
\label{Sec:AppendixB}
Cosine beams are generated by two waves with opposite directions. Hence, in the wave representation, the two waves have in-plane $k$-components of opposite sign, $+k_x=+k\sin \psi$, $-k_x=-k\sin \psi$. The angle $\psi$ is the angle formed by each ray with respect to the $z$-axis. The input phase can be written concisely as
\begin{align}
     \phi(x)=k\sin{\theta}x-k\beta_x |x|,
    \label{Eq:EqB1}
\end{align}
where $\theta$ is the steering angle and $\beta_x\equiv\sin\psi$. With simple inspection of Fig.\,\ref{fig:fig02}, the angle $\psi$ is associated with the ULA size $N_xd_x$ and the convergence distance $z_\mathrm{max}$ as
\begin{align}
     \tan \psi = \frac{N_xd_x}{2z_\mathrm{max}}.
    \label{Eq:EqB2}
\end{align}
Using the trigonometric identity
\begin{align}
     \sin{\psi}=\frac{\tan \psi}{\sqrt{1+\tan^2\psi}},
    \label{Eq:EqB3}
\end{align}
we find that
\begin{align}
     \beta_x=\frac{\frac{N_xd_x}{2z_\mathrm{max}}}{\sqrt{1+\left( \frac{N_xd_x}{2z_\mathrm{max}}\right)^2}}\approx \frac{N_xd_x}{2z_\mathrm{max}},
    \label{Eq:EqB4}
\end{align}
for $z_\mathrm{max}>N_xd_x$.
\subsection{Correlations of steering vectors at the Tx}
\label{Sec:AppendixC}
The correlation of two beams at the transmitter is written as
\begin{align}
    C=\frac{\iint_{-\infty}^{+\infty} E_1(x,y;0)E_2^*(x,y;0)dxdy}{\sqrt{\iint_{-\infty}^{+\infty} |E_1(x,y;0)|^2dxdy}\sqrt{\iint_{-\infty}^{+\infty} |E_1(x,y;0)|^2dxdy}},
    \label{Eq:EqC1}
\end{align}
where $E_1(x,y;0)=A_1(x,y) e^{j\phi_1(x,y)}$ and $E_2(x,y;0)=A_2(x,y) e^{j\phi_2(x,y)}$. 

For uniform amplitude $A_1(x,y)=A_2(x,y)=1$, and the correlation takes the simpler form
\begin{align}
    C=\frac{\int_{-D_x/2}^{+D_x/2} \int_{-D_y/2}^{+D_y/2} e^{j\phi_1(x,y)} e^{j\phi_2^*(x,y)}dxdy}{D_xD_y},
    \label{Eq:EqC2}
\end{align}
where the integration has been limited within the Tx area, which has size $D_x \times D_y$, with $D_x = N_xd_x$, $D_y = N_yd_y$.
In its discrete version, it becomes
\begin{align}
    C=\frac{1}{N_xN_y} \sum_{n_x}\sum_{n_y} e^{j\phi_1(n_x d_x, n_y d_y)} e^{j\phi_2^*(n_x d_x,n_y d_y)},
    \label{Eq:EqC3}
\end{align}
or, in terms of the corresponding steering vectors, $\textbf{a}_{1}$, $\textbf{a}_{2}$
\begin{align}
    C=\textbf{a}_{1}^H \textbf{a}_{2}.
    \label{Eq:EqC4}
\end{align}
\bibliographystyle{IEEEtran}
\bibliography{IEEEabrv,main}
\end{document}